\documentclass[prb,twocolumn,superscriptaddress]{revtex4}

\usepackage{graphicx}
\usepackage[centertags]{amsmath}
\usepackage{color}

\newcommand{\cm}{\ensuremath{\,\mbox{cm}^{-1}}}
\newcommand{\K}{\ensuremath{\,\mbox{K}}}

\newcommand{\Tn}{T$_{N}$}
\definecolor{gray}{rgb}{0.75,0.75,0.75}

\begin{document}
\title{Terahertz and infrared spectroscopic evidence of phonon-paramagnon
coupling\\
 in hexagonal piezomagnetic YMnO$_{3}$}

\author{C.~Kadlec}
\affiliation{Institute of Physics ASCR, Na Slovance~2, 182 21
Prague~8, Czech Republic}
\author{V.~Goian} \affiliation{Institute of Physics ASCR, Na Slovance~2, 182 21 Prague~8, Czech
Republic}
\author{K.\,Z.~Rushchanskii}
\affiliation{Peter Gr\"{u}nberg Institut, Forschungszentrum
J\"{u}lich GmbH, 52425 J\"{u}lich and JARA-FIT, Germany}
\author{P.~Ku\v{z}el}
\affiliation{Institute of Physics ASCR, Na Slovance~2, 182 21
Prague~8, Czech Republic}
\author{M.~Le\v{z}ai\'{c}}
\affiliation{Peter Gr\"{u}nberg Institut, Forschungszentrum
J\"{u}lich GmbH, 52425 J\"{u}lich and JARA-FIT, Germany}
\author{K.~Kohn}
\affiliation {Waseda University, Department of Physics, Tokyo
169-8555, Japan}
\author{R.\,V.~Pisarev}
\affiliation {Ioffe Physical-Technical Institute, Russian Academy of
Sciences, 194021 St. Petersburg, Russia}
\author{S.~Kamba}\email{kamba@fzu.cz}
\affiliation{Institute of Physics ASCR, Na Slovance~2, 182 21
Prague~8, Czech Republic}

\date{\today}

\begin{abstract}
Terahertz and far-infrared electric and magnetic responses of hexagonal piezomagnetic
YMnO$_{3}$ single crystals are investigated. Antiferromagnetic resonance is observed in
the spectra of magnetic permeability $\mu_{a}$ [\textbf{H}$\,(\omega)$ oriented within
the hexagonal plane] below the N\'{e}el temperature $T_{N}$. This excitation softens from
41 to 32\cm\, on heating and finally disappears above $T_{N}$. An additional weak and
heavily-damped excitation is seen in the spectra of complex dielectric permittivity
$\varepsilon_{c}$ within the same frequency range. This excitation contributes to the
dielectric spectra in both antiferromagnetic and paramagnetic phases. Its oscillator
strength significantly increases on heating towards room temperature thus providing
evidence of piezomagnetic or higher-order couplings to polar phonons. Other
heavily-damped dielectric excitations are detected near 100\cm\, in the paramagnetic
phase in both $\varepsilon_{c}$ and $\varepsilon_{a}$ spectra and they exhibit similar
temperature behavior. These excitations appearing in the frequency range of magnon
branches well below polar phonons could remind electromagnons; however, their temperature
dependence is quite different. We have used density functional theory for calculating
phonon dispersion branches in the whole Brillouin zone. A detailed analysis of these
results and of previously published magnon dispersion branches brought us to the
conclusion that the observed absorption bands stem from phonon-phonon and
phonon-paramagnon differential absorption processes. The latter is enabled by a strong
short-range in-plane spin correlations in the paramagnetic phase.

\end{abstract}


\maketitle

\section{Introduction}

Spin waves (magnons) in magnetically ordered materials can be excited by the magnetic
component \textbf{H}$(\omega)$ of the electromagnetic radiation, giving rise to a
resonant dispersion of magnetic permeability in the microwave or terahertz (THz)
frequency region. Recently, new coupled spin--lattice excitations named electromagnons
have been discovered in multiferroics, where the magnetic order coexists with the
ferroelectric one.\cite{pimenov06,sushkov07} Electromagnons are excited by the electric
component \textbf{E}$(\omega)$ of the electromagnetic radiation, therefore they can be
detected in the THz dielectric permittivity spectra. Though they were theoretically
predicted in 1970,\cite{chupis70} the first experimental confirmation appeared as late as
in 2006.\cite{pimenov06} These excitations were mainly investigated in the rare earth
(\emph{R}) orthorhombic manganites \emph{R}MnO$_3$ and \emph{R}Mn$_2$O$_5$ (for reviews
see e.g. Refs.\ \onlinecite{pimenov08,kida09,shuvaev11}), and in
hexaferrites.\cite{kida09b}

Multiferroics can be roughly divided into two
groups.\cite{khomskii06,khomskii09,lotter09} In the so-called type-I
multiferroics the ferroelectric (FE) order takes place both above
and below the magnetic ordering temperature and the spontaneous
polarization is large. However, the coupling between magnetic and
electric order parameters is weak.

A general feature of type-II multiferroic materials is that the ferroelectric phase is
induced by magnetic ordering characterized by a particular type of incommensurate spiral
magnetic structure. In this case the magnetically-induced polarization is by several
orders of magnitude smaller than in type-I multiferroics. However the coupling between
electric and magnetic subsystems is large and giant magnetoelectric effects are observed.
The magnon dispersion branch in the incommensurate phase exhibits a minimum at the wave
vector \textbf{q$_{m}$} corresponding to the modulation vector of the ordered spins. In
contrast to the magnetic resonance (magnon at \textbf{q}$\approx$0) characterized by
sharp spectral features, the electromagnons manifest themselves as very broad spectral
bands because their activation in the dielectric spectra is closely related to the high
density of states close to the extrema of the magnon dispersion branches. Since the
probing THz radiation has a long wavelength (i.e. the wave vector \textbf{q}$\approx$0),
the electromagnons cannot be excited by a resonant single-photon absorption due to the
wave vector conservation law; in this sense polar phonons should be involved in the
interaction process.

An experimentally observed low-frequency electromagnon in type-II multiferroics was found
to be related to the spin waves near the magnetic Brillouin zone (BZ) center with
\textbf{q}=\textbf{q$_{m}$} \cite{pimenov08,lee09} or to those with
\textbf{q}=\textbf{q}$_\textrm{BZE}$-2\textbf{q$_{m}$}; \cite{rovillain11} here
\textbf{q}$_\textrm{BZE}$ stands for the wave vector at the BZ edge. In both cases the
low-frequency electromagnon has a similar frequency as the magnon with q=0 which is
expected because all these excitations are related to the same magnon branch. A
high-frequency electromagnon corresponds to an excitation of the BZ-edge magnons
(\textbf{q}=\textbf{q}$_\textrm{BZE}$) which can induce a quasi-uniform modulation
(\textbf{q}$\approx$0) of the local electric dipole moment.\cite{lee09,valdes09} As for
mechanisms of the electromagnon excitations, some researchers claim that the
low-frequency electromagnons are activated by the inverse Dzyaloshinskii-Moriya
mechanism, while the high-frequency one by the Heisenberg exchange
coupling.\cite{kida09,shuvaev11} Other authors believe that both types of electromagnons
can be explained by the Heisenberg exchange coupling.\cite{mochizuki10}

Formerly, it was assumed that the electromagnons can be activated only in type-II
multiferroics due to the large magnetoelectric coupling. Nevertheless, electromagnons
were recently observed also in BiFeO$_{3}$, \cite{cazayous08,komandin10,talbayev11} which
is the most prominent type-I multiferroic with a rather weak magnetoelectric coupling. In
this context, THz dielectric spectra of multiferroics may shed new light on the nature of
magnetoelectric coupling.

Hexagonal manganites \emph{R}MnO$_3$ belong to the type-I multiferroics. In particular,
the hexagonal YMnO$_{3}$ is ferroelectric below $\approx$1250\K\cite{gibbs11} and the
antiferromagnetic (AFM) ordering sets only below T$_N\approx70$\K.\cite{bertaut63,
chatterji07} The magnetic symmetry is P\underline{6}$_3$c\underline{m} \cite{fiebig00}
and therefore the linear magnetoelectric coupling is forbidden. However, piezomagnetic
and magnetoelastic couplings, and higher-order magnetoelectric couplings are
allowed.\cite{Birss,fiebig02,goltsev03} The piezomagnetic coupling is characterized by a
bilinear interaction between the magnetic order parameter and strain, in contrast to the
magnetoelastic coupling which is proportional to the product of squared order parameter
and strain.\cite{Birss,Landau} By using the method of optical second harmonic
generation,\cite{Fiebig-JOSAB} the piezomagnetic coupling was observed owing to the
interaction between AFM and FE domain walls in YMnO$_{3}$.\cite{fiebig02,goltsev03}
Switching of the FE polarization triggers a reversal of the AFM order
parameter.\cite{fiebig02,goltsev03,choi10} Higher order magnetoelectric coupling in
YMnO$_{3}$ has been observed in several works. Exceptionally large atomic displacements
at $T_N$ were observed in structural studies and they demonstrate unusually strong
magnetoelastic coupling.\cite{lee08} The large spin--polar-phonon coupling manifests
itself by a decrease of the low-frequency permittivity\cite{aikawa05} near $T_{N}$ which
is probably caused by anomalous hardening of several infrared-active
phonons\cite{zaghrioui08}. Similar phonon anomalies were observed near $T_{N}$ also in
the Raman spectra.\cite{fukumura07} Ultrasound measurements on a single crystal of the
hexagonal YMnO$_{3}$ showed anomalous behavior of the elastic moduli $C_{11}$ and
$C_{66}$ due to a strong coupling of the lattice with the in-plane exchange
interaction.\cite{poirier07}

The AFM resonance in hexagonal YMnO$_{3}$ crystal was first observed and briefly (without
any figures) reported in Ref.\ \onlinecite{penney69}. More detailed THz studies of
YMnO$_{3}$ ceramics were recently published in Ref.\ \onlinecite{goian10}. The AFM
resonance lies near 43\cm\,at 4\K\, and its frequency softens on heating towards $T_N$,
where it disappears.\cite{penney69,goian10} Three magnon branches were discovered below
$T_N$ using inelastic neutron scattering (INS).\cite{sato03,petit07,chatterji07} Two of
them are degenerated near the BZ center and their frequencies correspond to the above
mentioned AFM resonance. Moreover, a possible existence of magnons and short-range
correlations between spins at Mn sites in paramagnetic phase were indicated by
INS.\cite{park03,roessli05,demmel07} The magnetoelastic coupling manifests itself also by
a strong mixing of magnons with acoustic phonons; this leads to a gap in the transverse
acoustic (TA) phonon branch occurring at the frequencies and wave vectors where the
uncoupled magnon and TA branches would intersect.\cite{petit07} Recent polarized INS
measurements revealed that the excitation detected at liquid helium temperatures near
43\cm\, has a mixed character of magnetic spin wave and lattice
vibration,\cite{pailhes09} i.e. its contribution to both the magnetic permeability and
the dielectric permittivity is possible.

\begin{figure*}[ht!!]
    \includegraphics[width=17cm]{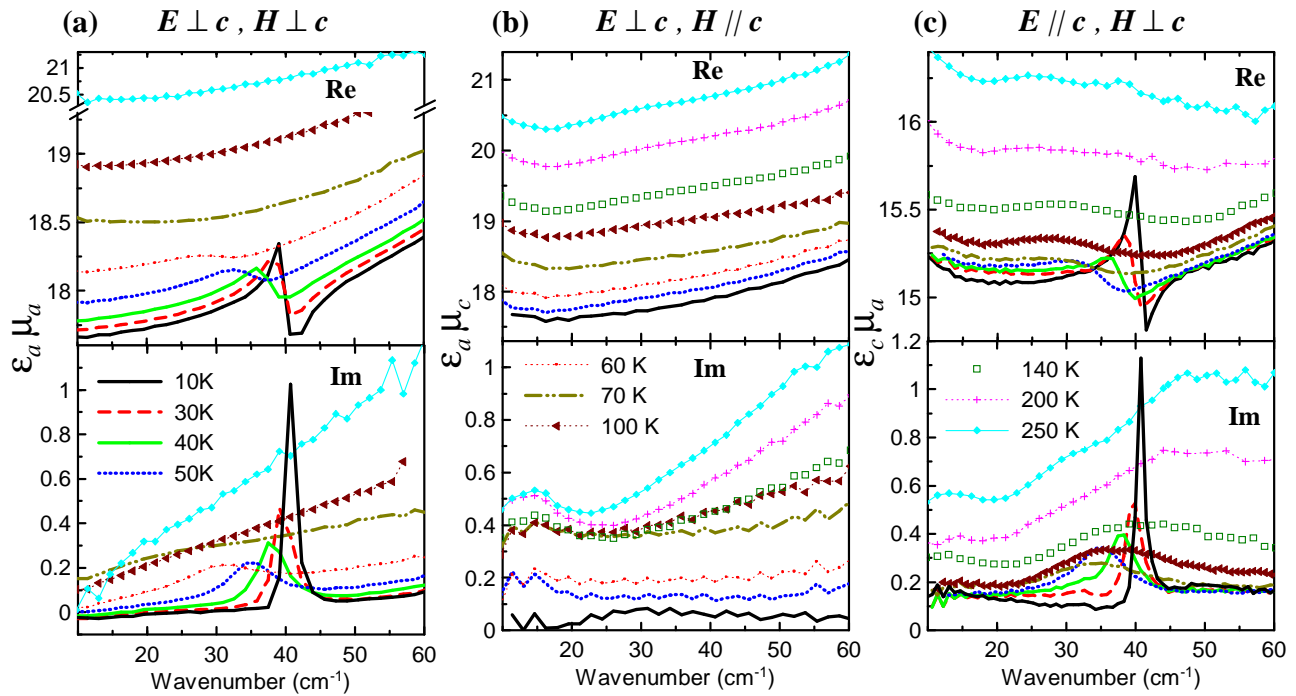}
    \caption{(Color online) Complex THz spectra of YMnO$_{3}$ taken at various temperatures.
    The polarization of the THz
    beam is indicated above the plots. The resonance feature near $\sim$40\cm\,corresponds to the
    doubly-degenerated AFM
    mode contributing to the magnetic permeability $\mu_a$ spectra.}
    \label{Fig1}
\end{figure*}

The reported piezomagnetic, magnetoelastic, and higher-order magnetoelectric couplings in
optical, acoustic and mainly INS data stimulated our spectroscopic study of hexagonal
single crystals of YMnO$_{3}$. In this paper, we present results on far-infrared (FIR)
and THz polarized spectra in this material emphasizing interaction between magnetic,
electric and phonon subsystems. We demonstrate that strongly underdamped AFM resonance
observed near $\approx$ 40\cm\, contributes only to the magnetic permeability spectra
below T$_{N}$. An additional broad and weak absorption band was observed in the same
frequency range in the dielectric spectra both below and above T$_{N}$. In contrast to
electromagnons which are typically observed only below 50\,K, the oscillator strength of
this excitation significantly increases on heating when room temperature is approached.
This indicates that the feature must be related to the occupation number of magnons
and/or phonons. Additional absorption band with similar temperature behavior was observed
also near 100\cm. We will show that both these excitations can be explained by
differential multiphonon and magnon-phonon processes.

\section{Experimental details}

The experiments were performed using a Fourier-transform infrared (FTIR) spectrometer
Bruker IFS\,113v and a custom-made THz time-domain spectrometer.\cite{kuzel10} In both
experiments, Optistat CF cryostats (Oxford Instruments) with polyethylene (FIR) or Mylar
(THz) windows were used for measurements between 10 and 300\K. Helium-cooled bolometer
operating at 1.6\,K was used as a detector in the FTIR spectrometer. Principles of THz
time-domain spectroscopy are explained in Ref.\ \onlinecite{dexheimer08}. The output of a
femtosecond Ti:sapphire laser oscillator (Coherent, Mira) excites an interdigitated
photoconducting switch TeraSED (Giga-Optics) to generate linearly polarized broadband THz
probing pulses.
 A gated detection scheme based on an electro-optic sampling
with a 1\,mm-thick [110] ZnTe crystal permits to measure the time
profile of the electric field of the transmitted THz pulse (see Ref.
\onlinecite{kuzel10} for further details).

\begin{figure}
  \begin{center}
    \includegraphics[width=85mm]{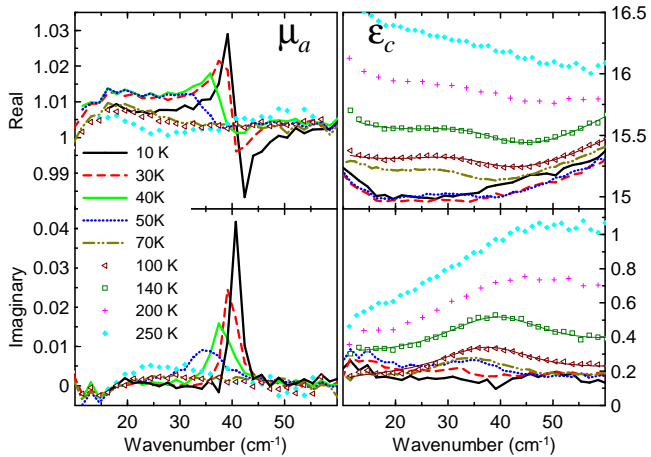}
  \end{center}
    \caption{(Color online) Temperature dependence of the complex permittivity $\varepsilon_c$
    and permeability
    $\mu_a$ spectra calculated from data plotted in Fig.~\ref{Fig1}. The solid $\varepsilon_c$ curves
    at 100 and 140\K\, result from the oscillator fit.}
    \label{Fig2}
\end{figure}

\begin{figure}
  \begin{center}
    \includegraphics[width=75mm]{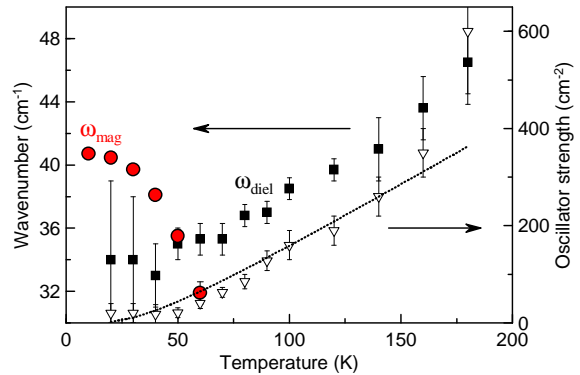}
  \end{center}
    \caption{(Color online) Temperature dependences of parameters of the resonances observed in
    magnetic $\mu_{a}$ and dielectric $\varepsilon_{c}$ spectra.
    Closed circles: frequency of the AFM resonance; Open squares and triangles: eigen-frequency $\omega_{diel1}$ and
    oscillator strength $\Delta\varepsilon\omega_{diel1}^{2}$,
    respectively, of the mode observed in the dielectric spectra in Fig.~\ref{Fig2}. The dotted line shows the
    population increase of an energy level at 66\,\cm\, following the Bose-Einstein statistics.}
    \label{Fig3}
\end{figure}

Hexagonal YMnO$_{3}$ single crystals were grown by the floating zone method.\cite{Kohn00}
Two crystal plates with lateral dimensions of $\sim4.5\times 5$ mm$^2$ and with the
{\itshape c}-axis oriented either in-plane or out-of-plane along its normal, were cut and
polished to obtain highly plane-parallel samples (within $\pm$ 1 $\mu$m) with thicknesses
of 1100 and 348 $\mu$m for each orientation, respectively. These crystal plates were
probed using the THz and FIR beam in all possible geometries:
\textbf{E}$\,(\omega)$$\perp$\textbf{c}, \textbf{H}$\,(\omega)$$\perp$\textbf{c};
\textbf{E}$\,(\omega)$$\perp$\textbf{c}, \textbf{H}$\,(\omega)$$\|$\textbf{c} and
\textbf{E}$\,(\omega)$$\|$\textbf{c}, \textbf{H}$\,(\omega)$$\perp$\textbf{c}. It enabled
us to get access to the complex spectra of the products $\varepsilon_{a} \mu_{a}$,
$\varepsilon_{a} \mu_{c}$, and $\varepsilon_{c} \mu_{a}$ as shown in Fig.~\ref{Fig1}(a),
(b), and (c), respectively.

\section{Results}

At low temperatures, the peak around 40 cm$^{-1}$ seen in the spectra of $\varepsilon_{a}
\mu_{a}$ and $\varepsilon_{c} \mu_{a}$ [Fig.~\ref{Fig1}(a, c)] but not in those of
$\varepsilon_{a} \mu_{c}$ [Fig.~\ref{Fig1}(b)] is definitely due to the AFM resonance as
it contributes only to the magnetic permeability $\mu_{a}$. The AFM resonance vanishes
above $T_{N}\sim70\,{\rm K}$. The data shown in Fig.~\ref{Fig1}(b) allow us to assume
that $\mu_{c}=1$ in the THz range. This is in agreement with the magnetic order of
YMnO$_{3}$ in the AFM phase: the spins are ordered in adjacent layers in the hexagonal
plane in such a way that the magnetic resonances are not expected to be excited with
\textbf{H}$\|$\textbf{c}. Based on this assumption, we are able to retrieve the complex
values of the permeability $\mu_{a}$ and of the permittivity $\varepsilon_{c}$ (see
Fig.~\ref{Fig2}).

The spectra of $\mu_{a}$ were fitted by a damped harmonic oscillator and the resulting
AFM resonance frequency is plotted in Fig.~\ref{Fig3}; a strong softening is observed
upon heating towards $T_{N}$. Similar temperature dependence was briefly published
earlier,\cite{penney69,goian10} with the magnon frequency higher by approximately 2\cm.

Besides the sharp AFM resonance line in the low-temperature $\mu_{a}$ spectra one can
observe a broad dielectric absorption band around 40\cm\ in the $\varepsilon_{c}$
spectra. This feature is detected even above $T_N$, where its strength remarkably
increases with temperature. The presence of such a resonance in $\varepsilon_{c}$ is
qualitatively expected from a simple comparison of the raw data in Figs.~\ref{Fig1}(a)
and (c). The accessible spectral range of the THz measurements for our sample is limited
to $\sim$ 60\cm, therefore we have performed also FTIR transmission (up to 100\cm) and
reflectivity (up to 650\cm) measurements for all polarizations.

\begin{figure}
  \begin{center}
    \includegraphics[width=86mm]{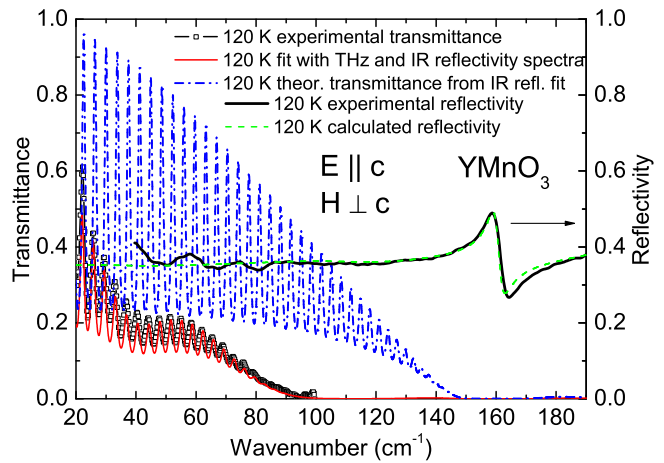}
  \end{center}
    \caption{(Color online) Example of the experimental FTIR transmittance and reflectivity spectra of 348\,$\mu$m
    thick YMnO$_3$ crystal with polarization
    \textbf{E}$\|$\textbf{c} obtained at 120\,K. Dashed-dotted blue line: theoretical transmittance spectrum
    obtained from
    parameters of the FTIR reflectivity fit (without considering
    modes observed by THz spectroscopy); solid red line: simultaneous fit of the FTIR transmission and THz spectra.
    Dashed green line is the result of a fit of the reflectivity using the parameters obtained from the fit of
    FIR and THz transmittance. One can see that the reflectivity spectrum is not sensitive enough to detect
    the weak broad modes near 40 and 100\,\cm. Oscillations in the experimental reflectivity spectrum observed
    below 80\,\cm\, are caused by the diffraction of FIR beam on a small sample.}
    \label{Fig4}
\end{figure}

\begin{figure}
  \begin{center}
    \includegraphics[width=85mm]{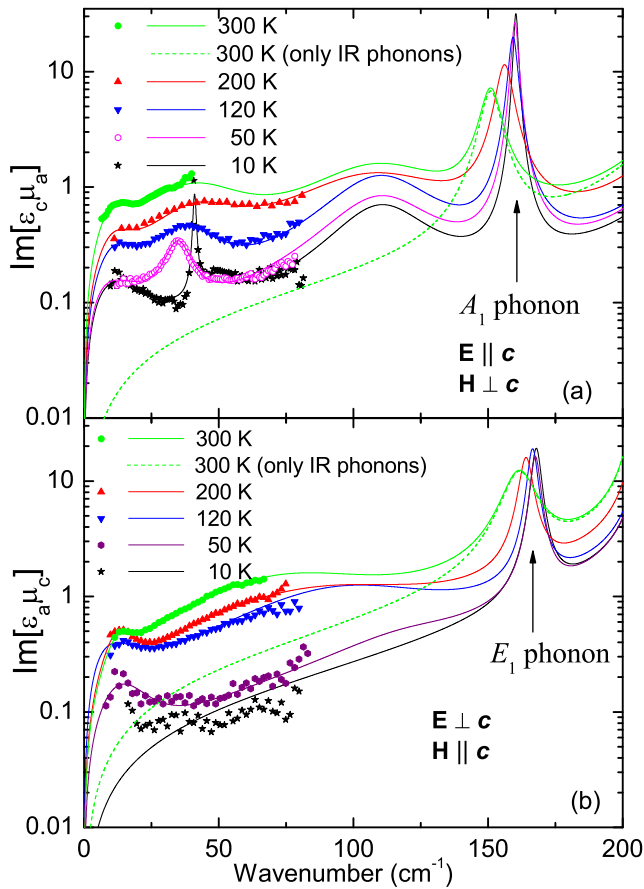}
  \end{center}
    \caption{(Color online)  The measured THz loss spectra of YMnO$_3$ (symbols) and those
    obtained from the fits of FTIR transmittance and reflectivity spectra. Below T$_{N}$=70\K,
    the spectra  correspond
    to imaginary parts of the permittivity-permeability product. Above \Tn, the spectra correspond
    to the dielectric losses.
    Polarizations of electric and magnetic components of IR or THz beams are indicated.
    Dashed lines are the fits of
    room-temperature FTIR reflectivity spectra without taking into account the IR and THz
    transmittance spectra.
    The marked peaks above 150\cm\ are due to phonons; the origin of lower frequency absorption
    bands is discussed in the text.}
    \label{Fig5}
\end{figure}

An example of FTIR experimental transmittance and reflectivity spectra obtained at 120\K\
and their various fits are shown in Fig.~\ref{Fig4}. Regular oscillations observed in the
transmittance spectrum are due to Fabry-P\'{e}rot interferences in the plane-parallel
sample; a weak minimum near 40\cm\, corresponds to the broad absorption band detected in
the THz dielectric spectra (see Fig.~\ref{Fig2}). According to Ref.\
\onlinecite{zaghrioui08} as well as according to our FTIR reflectivity (see e.g.
Fig.~\ref{Fig4}), the lowest frequency polar phonons lie above 150\cm\, in both polarized
\textbf{E}$\|$\textbf{c} and \textbf{E}$\perp$\textbf{c} spectra. Nevertheless, our
simultaneous fits of the THz complex permittivity and FTIR transmittance and reflectivity
data reveal several additional modes below these phonon frequencies. The relevant spectra
are plotted in Fig.~\ref{Fig5}. Besides the sharp magnon line at 40\cm\; three other
broad modes at roughly 10, 40 and 100\cm\, were used in the fitting procedure in order to
account for the measured shape of the \textbf{E}$\|$\textbf{c} spectra at 10\K\, (see
Fig.~\ref{Fig5}a). The additional modes remain in the spectra up to room temperature and
their strength increases on heating. Also in \textbf{E}$\perp$\textbf{c} polarized
spectra, two broad modes observed near 10 and 90\cm\ were used for the fits above 50\K.

The feature observed near 10\cm\, in both polarized spectra could be related to
low-frequency magnons\cite{sato03} (cf.\ the low-frequency magnon branches shown in
Fig.~\ref{fig:dispersion}). However, the sensitivity and accuracy of our THz spectra
below 20\cm\ is limited; therefore we cannot exclude that it is only an artifact. For
this reason we will not speculate about the origin of this excitation. All other modes
appearing below 150 \cm\ are clearly observed in the THz and/or FTIR transmittance
spectra while the FTIR reflectivity measurements are not sensitive enough to detect and
resolve these weak and broad spectral features (see Fig.~\ref{Fig4}). Their origin will
be discussed in the next section.

\begin{figure}[htbp]
\includegraphics[width=7cm]{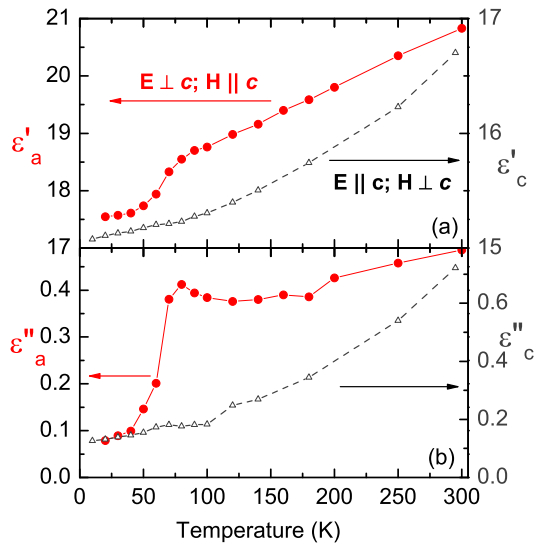}
\caption{(Color online) Temperature dependence of the (a) permittivity and (b) dielectric
loss measured at 20\cm\, with polarization \textbf{E}$\perp$\textbf{c} (red solid lines)
and \textbf{E}$\parallel$\textbf{c} (black dashed lines).} \label{fig:eps-T}
\end{figure}

The temperature dependence of the sub-THz complex dielectric permittivity
$\varepsilon_{a}$ plotted in Fig.~\ref{fig:eps-T} for 20\cm\, exhibits a pronounced drop
below $T_{N}$. Such anomaly is a typical feature of large spin-phonon coupling which
occurs only in hexagonal planes of YMnO$_{3}$, where the spins are ordered. For that
reason the anomaly is not observed in $\varepsilon_{c}$(\textit{T}). The AFM phase
transition is accompanied by unusually large atomic displacements, which were detected by
neutron diffraction;\cite{lee08} for this reason the phonon frequencies change below
$T_{N}$. The decrease in $\varepsilon'_{a}$ and $\varepsilon''_{a}$ is mainly caused by
hardening of the $E_{1}$ symmetry polar mode seen near 250\cm\, in the IR reflectivity
spectra with polarization \textbf{E}$\perp$\textbf{c}.\cite{zaghrioui08} Fits of our IR
reflectivity spectra show that the mode near 250\,\cm\ hardens from 246\, \cm\, (at 300
K) to 256\,\cm\, (at10\,K) and therefore its dielectric contribution $\Delta\varepsilon$
is reduced from 9.1 (300\,K) to 7.6 (10\,K). This decrease of $\Delta\varepsilon$ is
mainly responsible for the change of the permittivity $\varepsilon'_{a}$(T) seen in
Fig.~\ref{fig:eps-T}. Hardening of other modes brings a minor contribution to the
decrease of $\varepsilon'_{a}$(T) on cooling. Similar temperature dependence of
$\varepsilon'_{a}$ was observed also in the radio-frequency region\cite{aikawa05}
providing evidence of the absence of dielectric dispersion below 100 GHz. Gradual
decrease of $\varepsilon'_{a}$ and $\varepsilon'_{c}$ on cooling from 300 to 100\K\, is a
usual behavior caused by a small phonon stiffening as a consequence of thermal
contraction.

\begin{figure}[htbp]
\includegraphics[width=8.3cm]{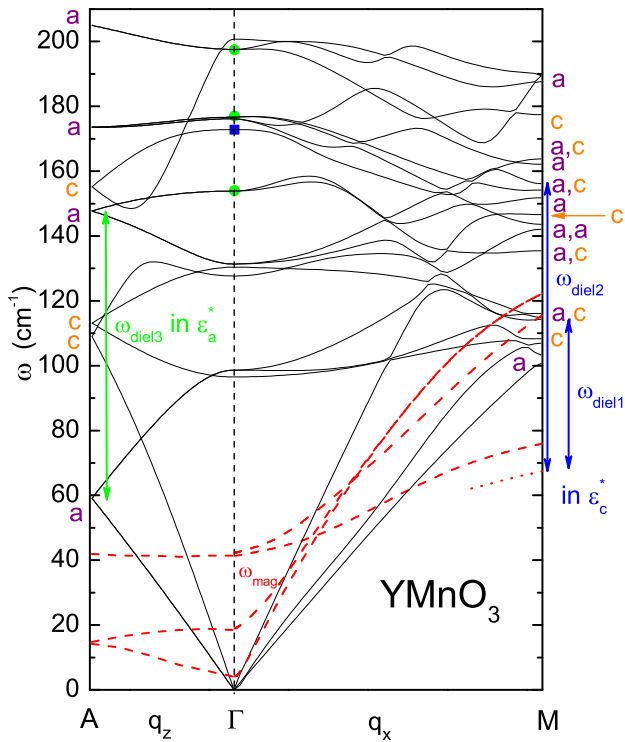}
\caption{(Color online) Dispersion branches of phonons (theoretical;
black solid lines) and magnons (experimental\cite{sato03} at 7\K;
red dashed lines). The red-dotted line indicates the presumable
dispersion of the paramagnon near the M-point. The symbols shown at
the BZ edges indicate the polarization of the phonons at the BZ
boundary: \textit{a} and \textit{c} stand for phonons polarized
within the hexagonal plane and in the perpendicular direction,
respectively. In the $\Gamma$-point, the \textit{E$_{1}$}- and
\textit{A$_{1}$}-phonons observed experimentally\cite{zaghrioui08}
are marked by green and blue points, respectively; other modes are
silent. Blue arrows with assignment $\omega_{diel1}$ and
$\omega_{diel2}$ indicate phonon-paramagnon excitations observed in
the dielectric loss spectra of $\varepsilon''_c$. Green arrow marked
as $\omega_{diel3}$ indicates a broad multiphonon absorption
observed in the $\varepsilon''_a$ loss spectra (see
Fig.~\ref{Fig5}).} \label{fig:dispersion}
\end{figure}

\section{Discussion}

The question arises about the origin of the absorption bands appearing below phonon
resonances in Fig.~\ref{Fig5}. They are much weaker and significantly broader than those
of polar phonons and their strength increases when the temperature is increased, i.e.,
the strength is high in the paramagnetic phase. Their frequencies lying in the range of
40--100 \cm\, coincide with those of the magnon branch observed by INS at 7 K over the
BZ\cite{sato03} (see Fig.~\ref{fig:dispersion}). In the following text we discuss whether
these features can be related to the magnon dispersion branches.

Could a spin wave still exist in hexagonal YMnO$_{3}$ at room temperature? It is well
established that Mn spins exhibit a strong short-range correlation in hexagonal
YMnO$_{3}$ far above T$_{N}$. This was proved by an anomalous behavior of the
thermal conductivity,\cite{sharma04} elastic moduli,\cite{poirier07} as well as by
neutron scattering experiments.\cite{park03,roessli05,demmel07} Nevertheless, due to the
short-range correlation of the spins in the hexagonal plane of YMnO$_{3}$, one can expect
the existence of only short-wavelength paramagnons, i.e. magnons with large wavevectors
\textbf{\textit{q$_{x}$}} near the M-point of the BZ. A part of such a paramagnon branch
is schematically plotted in Fig.~\ref{fig:dispersion}. Note that its frequency is lower
than that of the magnon branch at 7\K, as the magnon frequency decreases by almost
10\cm\, on heating towards T$_{N}$ (see Fig.~\ref{Fig2}).

Electromagnons are excitations with frequencies close to those of spin waves, which, due
to specific couplings, are activated in the dielectric spectra. In perovskite manganites,
the parts of magnon branches exhibiting a high density of states are mainly involved in
these interactions (at BZ edge or close to the spin modulation wave
vector).\cite{valdes09} However, these electromagnons were observed only at very low
temperatures (typically less than 50\K). Their strength dramatically decreases on heating
and they usually disappear from the spectra at T$_{N}$ or close above
T$_{N}$.\cite{pimenov08,kida09,shuvaev11} This is in contradiction with our observations
in YMnO$_{3}$.

We came to the conclusion that the broad absorption bands we observe in the dielectric
spectra reflect excitations which must be coupled to phonons. Let us discuss in brief
which types of interaction between the magnetic subsystem and other degrees of freedom
might be expected on the basis of the point group crystallographic symmetry 6$mm$ and the
magnetic symmetry $\underline{6}m\underline{m}$.\cite{fiebig00} The magnetic order
parameter of YMnO$_3$ was analyzed in several publications and it was shown to transform
following B$_1$ ($\Gamma_4$) irreducible representation of the 6$mm$
group.\cite{Nedlin,Pashkevich,Sa,Koster} The $\underline{6}m\underline{m}$ symmetry
strictly forbids the linear magnetoelectric effect, i.e. bilinear terms
$\alpha_{ij}H_iE_j$ (where $H_i$ and $E_i$ are components of the magnetic and electric
field, respectively) are not allowed in the thermodynamic potential.\cite{Birss} However,
a higher order magnetoelectric effect (called sometimes the magnetodielectric effect),
accounted for by the $\beta_{ijk}H_iH_jE_k$ terms in the thermodynamic potential, is
allowed. This effect manifests itself in our measurements as a kink near $T_N$ in the
temperature dependence of $\varepsilon'_a$ (see Fig.\,\ref{fig:eps-T}).

The magnetic symmetry of YMnO$_3$ allows the piezomagnetic contribution to the
thermodynamic potential described by the terms $p_{ijk}H_{i}\sigma_{jk}$, where
$\sigma_{jk}$ is a stress component and $p_{ijk}$ denotes components of the piezomagnetic
tensor.\cite{Birss,goltsev03} We believe that this type of bilinear coupling must play an
important role in the interaction between the magnetic subsystem and the lattice.
Usually, the piezomagnetic effect is allowed thanks to the relativistic part of
spin-lattice and spin-spin interactions, provided the symmetry restrictions are
met.\cite{Landau} However, in YMnO$_3$ which is a noncollinear antiferromagnet the
exchange (Coulomb) interactions may be by several orders of magnitude stronger than the
relativistic ones and, therefore, they can be the origin of piezomagnetism.\cite{Vitebsk}
For example, extraordinary spin-phonon interactions were shown to contribute to the
thermal conductivity of YMnO$_3$ below $T_N$.\cite{sharma04} Higher order effects such as
$p_{ijkl}H_iH_j\sigma_{kl}$ are naturally also allowed in YMnO$_3$.

In order to provide a more quantitative explanation of the interaction between magnetic
subsystem and phonons, we calculated the phonon spectrum from first principles within the
spin-polarized local density approximation \cite{PhysRevB.23.5048}. We used projector
augmented-wave potentials as implemented in Vienna \textit{Ab Initio} Simulation Package
(VASP) \cite{VASP_Kresse:1993,VASP_Kresse:1996,Bloechl:1994, VASP_Kresse:1999}. The
following valence-electron configurations were considered: $4s^{2}4p^{6}5s^{2}4d^{1}$ for
Y, $3p^{6}4s^{2}3d^{5}$ for Mn, and $2s^{2}2p^{4}$ for oxygen. To account for the strong
electron correlation effects on the {\itshape d}-shells of Mn atoms, we used LDA+U
approach \cite{Anisimov_et_al:1997,Dudarev} with an on-site Coulomb parameter $U=8.0$~eV
and Hund's exchange $J_H=0.88$~eV as calculated in Ref.~\onlinecite{Medvedeva_2000}. The
spin-orbit interaction was not taken into account. We used an A-type antiferromagnetic
structure, where spins on Mn honeycomb layers are aligned ferromagnetically and the
layers with opposite spin-direction alternate along the $c$-axis.
\cite{Spaldin_NMat_2004,fennie05} A kinetic energy cutoff of 500~eV and a $4 \times 4
\times 2$ $\Gamma$-centered $k$-point mesh was used in the structural relaxation of the
unit cell, where the Hellman-Feynman forces were minimized to a value smaller than
0.5~meV/$\mathrm{\AA}$. Phonon calculations were performed on a $2 \times 2 \times 1$
$\Gamma$- centered $k$-point mesh, with a $2 \times 2 \times 2$ supercell within the
force-constant method. \cite{Kunc:1982, Alfe:2009} The Hellman-Feynman forces were
calculated for displacements of atoms of up to 0.04~$\mathrm{\AA}$. The dynamical matrix
for each $q$-point in the BZ was constructed by a Fourier transformation of the force
constants, calculated for the $\Gamma$-point and for the BZ boundaries. Phonon-mode
frequencies and atomic displacement patterns for each $q$-point were obtained as
eigenvalues and eigenvectors of the dynamical matrices. The result for directions
A-$\Gamma$-M and wavenumbers up to 200~cm$^{-1}$ are presented in
Fig.~\ref{fig:dispersion}.

As we have already pointed out, the absorption strength
significantly increases on heating. This is typical for difference
frequency absorption. Such process includes the annihilation of one
quasi-particle (phonon or magnon) with frequency $\omega_{1}$ and
the creation of another quasi-particle with a higher frequency
$\omega_{2}$. The dielectric resonance then occurs at frequency
$\omega_{diel}=\omega_{2}-\omega_{1}$. This process can involve
excitations from the whole BZ provided that the total wave vector is
conserved. The contribution of the parts of the dispersion branch
with the highest density of states is expected to dominate. The high
number of available states is found namely at the flat parts of the
bands close to the BZ boundaries, as it was observed, for example,
in MgO.\cite{komandin09}

Obviously such a process is strongly temperature dependent, as it is
related to the population of excitations with frequency
$\omega_{1}$, which follows the Bose-Einstein statistics. At low
temperatures the population of the levels which we study is close to
zero and the differential absorption then practically vanishes. It
becomes more probable when the energy level is thermally populated
at higher temperatures. This is in qualitative agreement with our
observations.

The differential transitions at the BZ boundary are possible only between phonons with
the same symmetry and if the total wave vector is conserved (i.e. the transition must be
 vertical in the wave-vector space). The broad absorption around
90\cm\, seen in $\varepsilon_{a}$ spectra (Fig.~\ref{Fig5}b) can be explained by
differential multiphonon absorption. Phonons near 60 and 150\cm\, at the A-point of BZ
are polarized in the hexagonal plane (marked as \textit{a} in Fig.~\ref{fig:dispersion})
and their difference gives the frequency $\omega_{diel3}$ = 90\cm\,, as observed.

However, the two bands seen in $\varepsilon_{c}$ spectra around $\omega_{diel1}$ = 40\cm\
and $\omega_{diel2}$ = 100\cm\ are impossible to explain by multiphonon absorption. The
frequency of the \textit{c}-polarized phonons at the BZ edge is higher than 100\cm. It
means that the population of such phonons should be much lower than that of the
\textit{a}-polarized phonon at 60 \cm. For this reason the strength of the differential
multiphonon absorption in the $\varepsilon_{c}$ spectra should be weaker than in
$\varepsilon_{a}$ spectra. Moreover, within such a hypothesis, a continuous absorption
band would be expected in the spectra due to the large number of \textit{c}-polarized
phonons at the M-point (see scheme in Fig.~\ref{fig:dispersion}). This is in
contradiction with the experimental results presented in Fig.~\ref{Fig5}.

We assume the existence of paramagnons near the M-point, and in this case a differential
paramagnon-phonon absorption with several maxima can be obtained. Moreover, because of
the similar Bose-Einstein factor for the paramagnon close to 70\cm\ and phonon near
60\cm\, at the A-point, the absorptions observed in both $\varepsilon_{a}$ and
$\varepsilon_{c}$ should have comparable strengths. This fits well with the experiment.
The frequency $\omega_{diel1}$ increases on heating (Fig.~\ref{Fig3}) presumably due to
the softening of the paramagnon branch with increasing temperature. The increase of the
oscillator strength $\Delta\varepsilon\omega_{diel1}^{2}$ of the mode observed in Fig. 3
is compatible with the temperature increase of the Bose-Einstein factor: this is
demonstrated by the dotted line which shows the expected population increase of an energy
level at 66\,\cm\, (i.e. the frequency of paramagnon at \textbf{q}$_\textrm{BZE}$).

\section{Conclusions}

The THz and FTIR transmission spectra of hexagonal YMnO$_{3}$ clearly revealed two kinds
of excitations of different nature, which exist below polar phonon frequencies. The sharp
AFM resonance band observed near 40\cm\ at low temperatures broadens upon heating and
disappears close to $T_{N}$. This resonance is the main contributor to the magnetic
permeability $\mu_a$. Additional broad excitations were observed in the frequency range
40--100 \cm\ in the dielectric permittivity spectra in both the AFM and paramagnetic
phases. Our theoretical explanation of the activation of these excitations in the THz
dielectric spectra is based on a two-particle differential processes schematically shown
in Fig.~\ref{fig:dispersion}. The resonance observed in $\varepsilon_{a}$ spectra is
caused by differential phonon absorption in the A-point of the BZ. The two broad
absorption bands in $\varepsilon_{c}$ spectra were described as differential
phonon-paramagnon processes. The absorption strength of these excitations in the THz
spectra increases on heating due to the growing population of paramagnons and phonons
with temperature. This is possible in the paramagnetic phase owing to strong short-range
spin correlations within hexagonal planes of YMnO$_{3}$. The processes we observe in
YMnO$_{3}$, where the linear magnetoelectric coupling is forbidden, are clearly different
from the one responsible for the appearance of electromagnons in multiferroics with
spin-induced ferroelectricity.\cite{pimenov06,sushkov07,pimenov08,kida09} The multiphonon
absorptions are allowed by symmetry in all dielectric systems, while paramagnon-phonon
absorptions can be expected only in paramagnetic systems with a strong short-range
magnetic order (e.g. in hexagonal manganites). Magnon-phonon absorption should be also
detectable in all magnetically ordered systems (FM, AFM, ferrimagnets etc.) with
relatively high critical temperatures. In such conditions the magnons at the Brillouin
zone edge may become sufficiently populated to allow multiparticle effects in the
spectra. This may stimulate further THz and FIR studies of other magnetically polarizable
systems.

{\bfseries Acknowledgements}

The authors thank M. Mostovoy for valuable discussions. This work was supported by the
Czech Science Foundation (Project No. 202/09/0682), by AVOZ10100520, and by the Young
Investigators Group Program of the Helmholtz Association (Contract VH-NG-409). The
contribution of Ph.D. student V.G. has been supported by projects 202/09/H041 and
SVV-2011-263303. R.V.P. acknowledges the support by the RFBR (Project No. 09-02-00070).
The support of the J\"{u}lich Supercomputing Center is gratefully acknowledged.


\begin{thebibliography}{10}

\bibitem{pimenov06} A. Pimenov, A.\,A. Mukhin, V.\,Yu. Ivanov, V.\,D. Travkin, A.\,M. Balbashov,
and A. Loidl, Nature Phys. \textbf{2}, 97-100 (2006).

\bibitem{sushkov07} A.\,B. Sushkov, R. V. Aguilar, S. Park, S.-W. Cheong, and H.\,D. Drew,
 Phys. Rev. Lett. \textbf{98}, 027202 (2007).

\bibitem{chupis70} V.\,G. Baryakhtar and I.\,E. Chupis, Sov. Phys.-Solid State
\textbf{11}, 2628 (1970).

\bibitem{pimenov08} A. Pimenov, A.\,M. Shuvaev, A.\,A. Mukhin, and A. Loidl,
 J. Phys.: Condens. Matter \textbf{20}, 434209 (2008).

\bibitem{kida09} N. Kida, Y. Takahashi, J.\,S. Lee, R. Shimano, Y. Yamasaki,
Y. Kaneko, S. Miyahara, N. Furukawa, T. Arima, and Y. Tokura, J.
Opt. Soc. Amer. B \textbf{26}, A35-A51 (2009).

\bibitem{shuvaev11} A.\,M. Shuvaev, A.\,A. Mukhin and A. Pimenov, J. Phys.: Condens. Matter
\textbf{23}, 113201 (2011).

\bibitem{kida09b} N. Kida, D. Okuyama, S. Ishiwata, Y. Taguchi, R. Shimano, K. Iwasa, T. Arima,
and Y. Tokura, Phys. Rev. B \textbf{80}, 220406(R) (2009).

\bibitem{khomskii06}D.\,I. Khomskii, J. Magn. Magn. Mater. \textbf{306}, 1
(2006).

\bibitem{khomskii09} D. Khomskii,  Physics \textbf{2}, 20 (2009).

\bibitem{lotter09}Th. Lottermoser, D. Meier, R.\,V. Pisarev, and M.
Fiebig, Phys. Rev. B \textbf{80}, 100101 (2009).

\bibitem{lee09} J.\,S. Lee, N. Kida, S. Miyahara, Y. Takahashi, Y. Yamasaki, R. Shimano, N. Furukawa,
and Y. Tokura, Phys. Rev. B \textbf{79}, 180403(R) (2009).

\bibitem{rovillain11} P. Rovillain, M. Cazayous, Y. Gallais, M-A. Measson, A. Sacuto, H. Sakata,
and M. Mochizuki, Phys. Rev. Lett. \textbf{107}, 027202 (2011).

\bibitem{valdes09} R. Vald\'{e}s Aguilar, M. Mostovoy, A.\,B. Sushkov, C.\,L. Zhang, Y.\,J.
Choi, S.-W. Cheong, and H.\,D. Drew, Phys. Rev. Lett. \textbf{102},
047203 (2009).

\bibitem{mochizuki10} M. Mochizuki, N. Furukawa, and N. Nagaosa, Phys. Rev.
Lett. \textbf{104}, 177206 (2010).

\bibitem{cazayous08}M. Cazayous, Y. Gallais, A. Sacuto, R. de Sousa, D. Lebeugle and D. Colson,
Phys. Rev. Lett. \textbf{101}, 037601 (2008).

\bibitem{komandin10} G. Komandin, V. Torgashev, A. Volkov, O. Porodinkov, I. Spektor, and A. Bush,
Phys. Sol. State \textbf{52}, 734 (2010).

\bibitem{talbayev11} D. Talbayev, S.\,A. Trugman, S. Lee, H.\,T. Yi, S.-W. Cheong, and A.\,J. Taylor,
Phys. Rev. B \textbf{83}, 094403 (2011).


\bibitem{gibbs11} A.\,S. Gibbs, K.\,S. Knight, and P. Lightfoot,
Phys. Rev. B \textbf{83}, 094111 (2011).

\bibitem{bertaut63} E. Bertaut and M. Mercier, Phys. Lett. \textbf{5}, 27 (1963).

\bibitem{chatterji07} T. Chatterji, S. Ghosh, A. Singh, L.\,P. Regnault, and
M. Rheinst\"{a}dter, Phys. Rev. B \textbf{76}, 144406 (2007).

\bibitem{fiebig00} M. Fiebig, D. Fr\"{o}hlich, K. Kohn, St. Leute, Th. Lottermoser, V.\,V. Pavlov,
and R.\,V. Pisarev, Phys. Rev. Lett. \textbf{84}, 5620-5623 (2000).

\bibitem{Birss}R.\,R. Birss, \emph{Symmetry and Magnetism}, North-Holland,
1967.

\bibitem{fiebig02} M. Fiebig, Th. Lottermoser, D. Fr\"{o}hlich, A.\,V. Goltsev,
and R.\,V. Pisarev, Nature \textbf{419}, 818-820 (2002).

\bibitem{goltsev03} A.\,V. Goltsev, R. V. Pisarev, Th. Lottermoser, and M. Fiebig,
Phys. Rev. Lett. \textbf{90}, 177204 (2003).

\bibitem{Landau}L.\,D. Landau and E.\,M. Lifshitz, \emph{Electrodynamics
of  Continuos Media,} 2ed., Pergamon, 1984.

\bibitem{Fiebig-JOSAB}M. Fiebig, V.\,V. Pavlov, and R.\,V. Pisarev, J.
Opt. Soc. Amer. B \textbf{22}, 96 (2005).

\bibitem{choi10} T. Choi, Y. Horibe, H.\,T. Yi, Y.\,J. Choi,  Wu Weida, and  S.-W. Cheong,
 Nature Mat. \textbf{9}, 253-258 (2010).

\bibitem{lee08} S. Lee, A. Pirogov, M. Kang, K.-H. Jang,
M. Yonemura, T. Kamiyama, S.-W. Cheong, F. Gozzo, N. Shin, H.
Kimura, Y. Noda, and J.-G. Park, Nature \textbf{451}, 805-809
(2008).

\bibitem{aikawa05} Y. Aikawa, T. Katsufuji, T. Arima,
 and K. Kato,  Phys. Rev. B \textbf{71}, 184418 (2005).

\bibitem{zaghrioui08} M. Zaghrioui, V. Ta Phuoc, R.\,A.  Souza, and M. Gervais,  Phys.
Rev. B \textbf{78}, 184305 (2008).

\bibitem{fukumura07} H. Fukumura, S. Matsui, H. Harima, K. Kisoda, T. Takahashi, T. Yoshimura,
and N. Fujimura,  J. Phys.: Condens. Matter \textbf{19}, 365239
(2007).

\bibitem{poirier07} M. Poirier, F. Lalibert\'{e}, L. Pinsard, and A. Revcolevschi Phys. Rev. B
\textbf{76}, 174426 (2007).

\bibitem{penney69} T. Penney, P. Berger, and K. Kritiyakirana, J. Appl. Phys. \textbf{40},
1234-1235 (1969).

\bibitem{goian10} V. Goian, S. Kamba, C. Kadlec, D. Nuzhnyy, P.
Ku\v{z}el, J. Agostino Moreira, A. Almeida and P.\,B. Tavares, Phase
Transitions \textbf{83}, 931 (2010).

\bibitem{sato03} T.\,J. Sato, S.-H. Lee, T. Katsufuji, M. Masaki, S. Park, J.\,R.\,D. Copley,
and H. Takagi,  Phys. Rev. B \textbf{68}, 014432 (2003).

\bibitem{petit07} S. Petit, F. Moussa, M. Hennion, S. Pailh\`{e}s,
L. Pinsard-Gaudart, and A. Ivanov, Phys. Rev. Lett. \textbf{99},
266604 (2007).

\bibitem{park03} J. Park, J.-G. Park, G.\,S. Jeon, H.-Y. Choi,
Ch. Lee, W. Jo, R. Bewley, K.\,A. McEwen, and T.\,G. Perring, Phys.
Rev. B \textbf{68}, 104426 (2003).

\bibitem{roessli05} B. Roessli, S.\,N. Gvasaliya, E. Pomjakushina, and K. Conder, JETP Letters \textbf{51}, 287 (2005).

\bibitem{demmel07} F. Demmel, T. Chatterji, Phys. Rev. B \textbf{76}, 212402 (2007).

\bibitem{pailhes09} S. Pailh\`{e}s, X. Fabr\`{e}ges, L.\,P. R\'{e}gnault,
L. Pinsard-Godart, I. Mirebeau, F. Moussa, M. Hennion, and S. Petit, Phys. Rev. B
\textbf{79}, 134409 (2009).


\bibitem{kuzel10} P. Ku\v zel, H. N\v emec, H., F. Kadlec, and C. Kadlec, Opt.
Express \textbf{18}, 15338 (2010).

\bibitem{dexheimer08} S.\,L. Dexheimer,  \emph{THz Spectroscopy: Principles and Applications} (Boca
Raton, FL: CRC Press).

\bibitem{Kohn00}H. Yamagichi, T. Fujita, T. Shinozaki, H. Sigie, and
K. Kohn, \emph{Ferrites: Proceedings of the Eighth International Conference on Ferrites
(ICF 8)}, Kyoto and Tokyo, Japan 2000.

\bibitem{sharma04} P.\,A. Sharma, J.\,S. Ahn, N. Hur, S. Park, S.\,B. Kim, S.
Lee, J.-G. Park, S. Guha, and S.-W. Cheong, Phys. Rev. Lett. \textbf{93}, 177202 (2004).



\bibitem{Nedlin} G.\,M. Nedlin, Sov. Phys. Solid St. \textbf{6}, 2165 (1965).

\bibitem{Pashkevich} Yu.\,G. Pashkevich, V.\,L. Sobolev, S.\,A.
Fedorov, and A.\,V. Eremenko, Phys. Rev. B \textbf{51}, 15898
(1995).

\bibitem{Sa}D. Sa, R. Valent\'{\i}, and C. Gros, Eur. Phys. Journ. B \textbf{14}, 301 (2000).

\bibitem{Koster}G.\,F. Koster, J.\,O. Dimmock, R.\,G. Wheeler, and
H. Statz, \emph{Properties of the Thirty Two Point Groups}, MIT,
1963.

\bibitem{Vitebsk}I.\,M. Vitebskii, N.\,M. Lavrinenko, and V.\,L.
Sobolev, J. Magn. Magn. Mater. \textbf{97}, 263 (1991).

\bibitem{fennie05} C.\,J. Fennie, and K.\,M. Rabe,
Phys. Rev. B \textbf{72} 100103(R) (2005).

\bibitem{PhysRevB.23.5048}J.\,P. Perdew, and A. Zunger, Phys. Rev. B \textbf{23}, 5048
(1981).

\bibitem{VASP_Kresse:1993} G. Kresse, and J. Hafner, Phys. Rev. B \textbf{47}, 558
(1993).

\bibitem{VASP_Kresse:1996} G. Kresse, and J. Furthm\"uller,
 Phys. Rev. B \textbf{54}, 11169 (1996).

\bibitem{Bloechl:1994} P.\,E. Bl\"ochl, Phys. Rev. B \textbf{50},
17953 (1994).

\bibitem{VASP_Kresse:1999} G. Kresse, and D. Joubert,  Phys. Rev. B \textbf{59}, 1758 (1999).

\bibitem{Anisimov_et_al:1997} V.\,I. Anisimov, F. Aryasetiawan, and A.\,I. Lichtenstein,
 J. Phys.: Condens. Matter \textbf{9}, 767 (1997).

\bibitem{Dudarev} S.\,L. Dudarev, G.\,A. Botton, S.\,Y. Savrasov,
C.\,J. Humphreys, and A.\,P. Sutton, Phys. Rev. B \textbf{57}, 1505
(1998).

\bibitem{Medvedeva_2000} J.\,E. Medvedeva, V.\,I. Anisimov, M.\,A. Korotin, O.\,N. Mryasov,
A.\,J. Freeman, J. Phys.: Condens. Matter \textbf{12}, 4947 (2000).

\bibitem{Spaldin_NMat_2004} B.\,B. van\,Aken, T.\,T.\,M. Palstra, A. Filippetti,
and N.\,A. Spaldin, Nat. Mater. \textbf{3}, 164 (2004).

\bibitem{Alfe:2009} D. Alf\`{e}, Comp. Phys. Commun. \textbf{180}, 2622 (2009).

\bibitem{Kunc:1982} K. Kunc and R.\,M. Martin,  Phys. Rev. Lett. \textbf{48},
406 (1982).

\bibitem{komandin09} G.\,A. Komandin, O.\,E. Porodinkov, I.\,E. Spector, and A.\,A. Volkov, Phys.
Sol. State \textbf{51}, 2045 (2009).





\end{thebibliography}
\end{document}